\documentclass[floatfix,aps,superscriptaddress,prd,amsmath,nofootinbib,preprintnumbers,twocolumn]{revtex4}

\usepackage{graphicx}
\usepackage{bm}
\usepackage{float}
\usepackage[
colorlinks=true,        
citecolor=blue,         
linkcolor=blue,         
urlcolor=blue           
]{hyperref}             

\newcommand{\nc}{\newcommand*}
\nc{\Om}{\Omega}
\nc{\ogw}{\Omega_{\mathrm{GW}}}
\nc{\rd}{\mathrm{d}}
\nc{\eg}{\textit{e.g.~}}
\nc{\red}[1]{\textcolor{red}{#1}}
\nc{\lvc}{LIGO/Virgo} 

\def\({\left(}
\def\){\right)}
\def\[{\left[}
\def\]{\right]}

\def\e{\begin{equation}}
\def\q{\end{equation}}
\def\m{\begin{eqnarray}}
\def\n{\end{eqnarray}}

\begin{document}

\title{Measuring the primordial gravitational waves from cosmic microwave background and stochastic gravitational wave background observations}

\author{Jun Li}
\email{lijun@qust.edu.cn}
\affiliation{School of Mathematics and Physics,
    Qingdao University of Science and Technology,
    Qingdao 266061, China}
\affiliation{CAS Key Laboratory of Theoretical Physics,
    Institute of Theoretical Physics, Chinese Academy of Sciences,
    Beijing 100190, China}

\author{Guang-Hai Guo}
\email{ghguo@qust.edu.cn}
\affiliation{School of Mathematics and Physics,
    Qingdao University of Science and Technology,
    Qingdao 266061, China}

\date{\today}

\begin{abstract}
We constrain the primordial gravitational waves from cosmic microwave background (CMB) and stochastic gravitational wave background (SGWB) observations. SGWB provides the latest way to explore the early universe and the cosmological evolution which can be reflected by primordial gravitational waves. We not only combine LIGO observations with CMB to measure primordial gravitational waves, but also forecast the potential abilities of the LISA detector and PTA projects. In the $\Lambda$CDM+$r$+$n_t$ model, the standard six parameters change slightly from SGWB observations. While the constraints on tensor-to-scalar ratio and tensor spectral index are improved obviously from SGWB observations. FAST projects have a significant impact on tensor-to-scalar ratio and tensor spectral index, namely $r<0.028$ and $n_t=-0.41^{+0.64}_{-0.96}$ at $95\%$ confidence level.

\end{abstract}

\maketitle

\section{introduction}
Since the discovery of temperature anisotropies in the cosmic microwave background, it has become significant to study the early universe and the cosmological evolution. The primordial fluctuations affect the temperature and polarization by scalar and tensor perturbation \cite{Cabella:2004mk,Kosowsky:1998mb,Zaldarriaga:1996xe,Ma:1995ey}. The CMB polarization can be decomposed into E-mode and B-mode. The B-mode component from tensor perturbation encodes the information of primordial gravitational waves \cite{Kamionkowski:1996zd,Kamionkowski:2015yta,BICEP2:2014owc,POLARBEAR:2014hgp}. Primordial gravitational waves cause the 
B-mode polarization of CMB and generate a stochastic gravitational waves background covering very wide frequency bands. The CMB power spectra indicate cosmological evolution information which can measure the primordial gravitational waves well. 

However,  the rapid developments of the gravitational wave observations inspire us to explore the new way to probe the early universe and the cosmological evolution which can be reflected by primordial gravitational waves. Gravitational waves provide the latest approach to constrain the primordial gravitational waves. After the Laser Interferometer Gravitational-wave Observatory (LIGO) Science Collaboration discovered the first direct detection of gravitational wave from the coalescence of binary black holes \cite{Abbott:2016blz}, many experiments are prepared to detect gravitational wave in a wide range of frequencies, such as the planned third-generation GW detectors: the Einstein Telescope \cite{Maggiore:2019uih} and the Cosmic Explorer \cite{Reitze:2019iox}. All of these observations are sensitive to stochastic gravitational waves background which contain the information of primordial gravitational waves. SGWB is a type of gravitational wave produced by an extremely large number of weak, independent, and unresolved sources. SGWB is from different sources, not only from the primordial gravitational waves. To achieve a better constraints on primordial gravitational waves, observations should be combined at different frequency bands. 

In this paper, we consider the observations from CMB and SGWB to obtain better constraints on primordial gravitational waves. The CMB constraints include Planck satellite\renewcommand{\thefootnote}{\Roman{footnote}}\footnote{Planck18: TTTEEE+lowE+lensing} \cite{Aghanim:2018eyx}, BICEP/Keck Observations through the 2018 Observing Season (BK18) \cite{BICEP:2021xfz}, Baryon Acoustic Oscillation (BAO)\renewcommand{\thefootnote}{\Roman{footnote}}\footnote{BAO: 6DF+MGS+DR12}  \cite{Beutler:2011hx,Ross:2014qpa,Alam:2016hwk}. The CMB polarization constrains the spectra in the very low frequency from $10^{-20}$ Hz to $10^{-15}$ Hz. The SGWB constraints include LIGO detector \cite{TheLIGOScientific:2016dpb, LIGOScientific:2019vic}, Laser Interferometer Space Antenna (LISA) detector \cite{Thrane:2013oya,Caprini:2015zlo} and two Pulsar timing array (PTA) projects \cite{Hellings:1983fr}, namely International Pulsar Timing Array (IPTA) \cite{Verbiest:2016vem} and Five-hundred-meter Aperture Spherical radio Telescope (FAST) \cite{Nan:2011um}. The data of Advanced LIGO are from the second observing run (O2) and first observing run (O1) in the high frequency from $20$ Hz to $1726$ Hz. The detect ability of LISA is in the frequency from $10^{-4}$ Hz to $1$ Hz. The detect ability of IPTA is in the low frequency from $1.58*10^{-9}$ Hz to $8.27*10^{-7}$ Hz and the detect ability of FAST is in the frequency from $6.34*10^{-10}$ Hz to $8.27*10^{-7}$ Hz. We combine the CMB data and the search results of SGWB to obtain better constraints on primordial gravitational waves.

\section{The detection ability for gravitational wave detectors}
The detection ability for gravitational wave detectors are characterized with signal-to-noise ratio (SNR).
For advanced LIGO detectors, SNR is given by \cite{Thrane:2013oya}
\e
\rho= \sqrt{2T} \[\int df \sum_{I, J}^{M} \frac{\Gamma_{IJ}^2(f)S_h^2(f)}{P_{nI}(f)P_{nJ}(f)}\]^{1/2},
\q
where $T$ is the observation time, $P_{nI}$ and $P_{nJ}$ are the auto power spectral densities for noise in detectors $I$ and $J$, $S_h(f)$ is the strain power spectral density of a stochastic gravitational wave background.
For an autocorrelation measurement in the LISA detector, SNR can be calculated by \cite{Thrane:2013oya,Caprini:2015zlo}
\e
\rho = \sqrt{T}\[\int df \Big(\frac{\Omega_{gw}}{\Omega_n}\Big)^2\]^{1/2},
\q
where $\Omega_n$ is related to the  strain noise power spectral density $S_n$.
For a PTA measurement, SNR can be obtained by
\e
\rho = \sqrt{2T} \(\sum_{I, J}^{M} \chi_{IJ}^2\)^{1/2} \[ \int \rd f \(\frac{\ogw(f)}{\Om_n(f) + \ogw(f)}\)^2 \]^{1/2},
\q
where $\chi_{IJ}$ is the Hellings and Downs coefficient for pulsars $I$ and $J$ \cite{Hellings:1983fr}.
We consider two PTA projects, namely IPTA \cite{Verbiest:2016vem} and FAST \cite{Nan:2011um}, and make the same assumptions for these PTAs as presented in a previous study \cite{Kuroda:2015owv}.

\begin{figure}
\centering
\includegraphics[width=7.8cm]{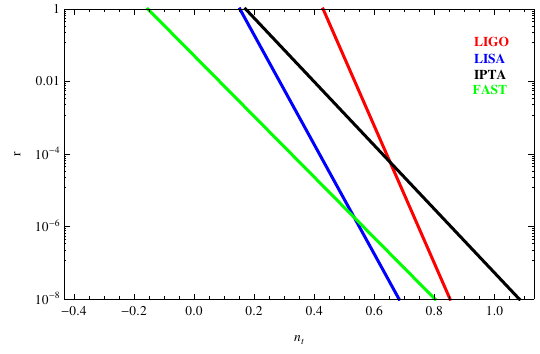}
\caption{The plot for parameters $n_t$ and $r$ from LIGO, LISA, IPTA and FAST. Here we assume the non-detection of stochastic gravitational wave background from future observations including LISA, IPTA and FAST. }
\label{fig:1}
\end{figure}

In order to characterize the spectral properties of stochastic gravitational wave background, the energy distribution in frequency is defined as follows
\e
\Omega_{gw}(f)=\frac{1}{\rho_c}\frac{d\rho_{gw}}{d\ln f}=\frac{2\pi^2}{3H_0^2}f^3 S_h(f). \label{gw1}
\q
Gravitational wave fractional energy density per logarithmic wavenumber interval today is given by \cite{Zhao:2013bba}
\m
\Omega_{gw}\simeq\frac{15}{16}\frac{\Omega_m^2 A_s r}{H_0^2\eta_0^4k_{\mathrm{eq}}^2}\Big(\frac{k}{k_*}\Big)^{n_t},\label{gw2}
\n
where $\Omega_m$ is the matter density, $H_0$ is the Hubble constant, $\eta_0=1.41\times 10^4$ Mpc denotes the conformal time today and $k_{\mathrm{eq}}=0.073\Omega_mh^2\ \mathrm{Mpc}^{-1}$ denotes the wavenumber when matter-radiation equality, $k_*=0.05$ Mpc$^{-1}$ denotes the pivot scale, $n_t$ is the tensor spectral index.
The parameter $r$, called tensor-to-scalar ratio, quantify the tensor amplitude $A_t$ compared to the scalar amplitude $A_s$ at the pivot scale
\e
r\equiv\frac{A_t}{A_s}.
\q
Combine the SNR definitions with Eq.~(\ref{gw1}), the sensitivity curves present the relation of $\Omega_{gw}(f)$ and frequency. According to Eq.~(\ref{gw2}), we regard $\Omega_{gw}$, $n_t$, $r$ and wavenumber as variations and fix other parameters to obtain functions of tensor-to-scalar ratio and tensor spectral index. If SGWB cannot be detected by gravitational wave observations, we obtain the upper limits on tensor-to-scalar ratio as the constraints from gravitational waves observations in Fig.~\ref{fig:1}. 
For LIGO detector, the real SGWB data can be found in Table 1 of Ref.~\cite{LIGOScientific:2019vic} and give the upper limit on tensor-to-scalar ratio
\m
r &<& 1.34*10^8*10^{-18.93n_t}.
\n
For LISA detector, the mock SGWB data give the upper limit on tensor-to-scalar ratio
\m
r &<& 2.00*10^2*10^{-15.10n_t}.
\n
For IPTA detector, the mock SGWB data give the upper limit on tensor-to-scalar ratio
\m
r &<& 3.20*10*10^{-8.78n_t}.
\n
For FAST detector, the mock SGWB data give the upper limit on tensor-to-scalar ratio
\m
r &<& 5.16*10^{-2}*10^{-8.37n_t}. \label{FAST}
\n
The upper bound on the tensor-to-scalar ratio from gravitational wave experiments depends from the value of the spectral index of the tensor power spectrum. If the tensor spectral index is negative, FAST projects give much tighter constraint than other observations. In this way, the primordial gravitational waves can be measured from CMB and SGWB observations. Our previous work has talked about measuring the tilt of primordial gravitational-wave power spectrum from observations \cite{Li:2019vlb}. We also discuss measuring the scalar induced gravitational waves from observations \cite{Li:2021uvn}. Some relevant works have also been reported in \cite{Huang:2015gka,Stewart:2007fu,Cabass:2015jwe,Lasky:2015lej,Yang:2019vni,Vagnozzi:2020gtf}.

\section{constraints on the primordial gravitational waves from CMB and SGWB observations}
\begin{figure}[tbh!]
\centering
\includegraphics[width=7cm]{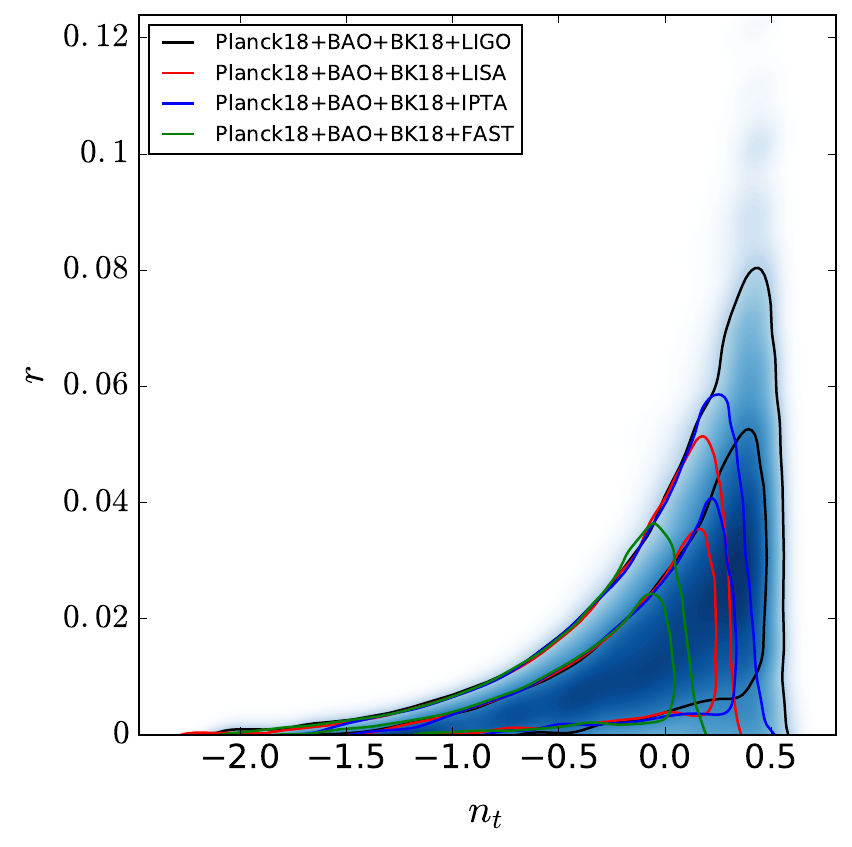}
\caption{The marginalized contour plot for parameters $n_t$ and $r$ in the $\Lambda$CDM+$r$+$n_t$ model at the $68\%$ and $95\%$ CL from the combinations of Planck18+BAO+BK18+LIGO, Planck18+BAO+BK18+LISA, Planck18+BAO+BK18+IPTA and Planck18+BAO+BK18+FAST, respectively.}
\label{fig:2}
\end{figure}

\begin{figure*}
\centering
\includegraphics[width=15.7cm]{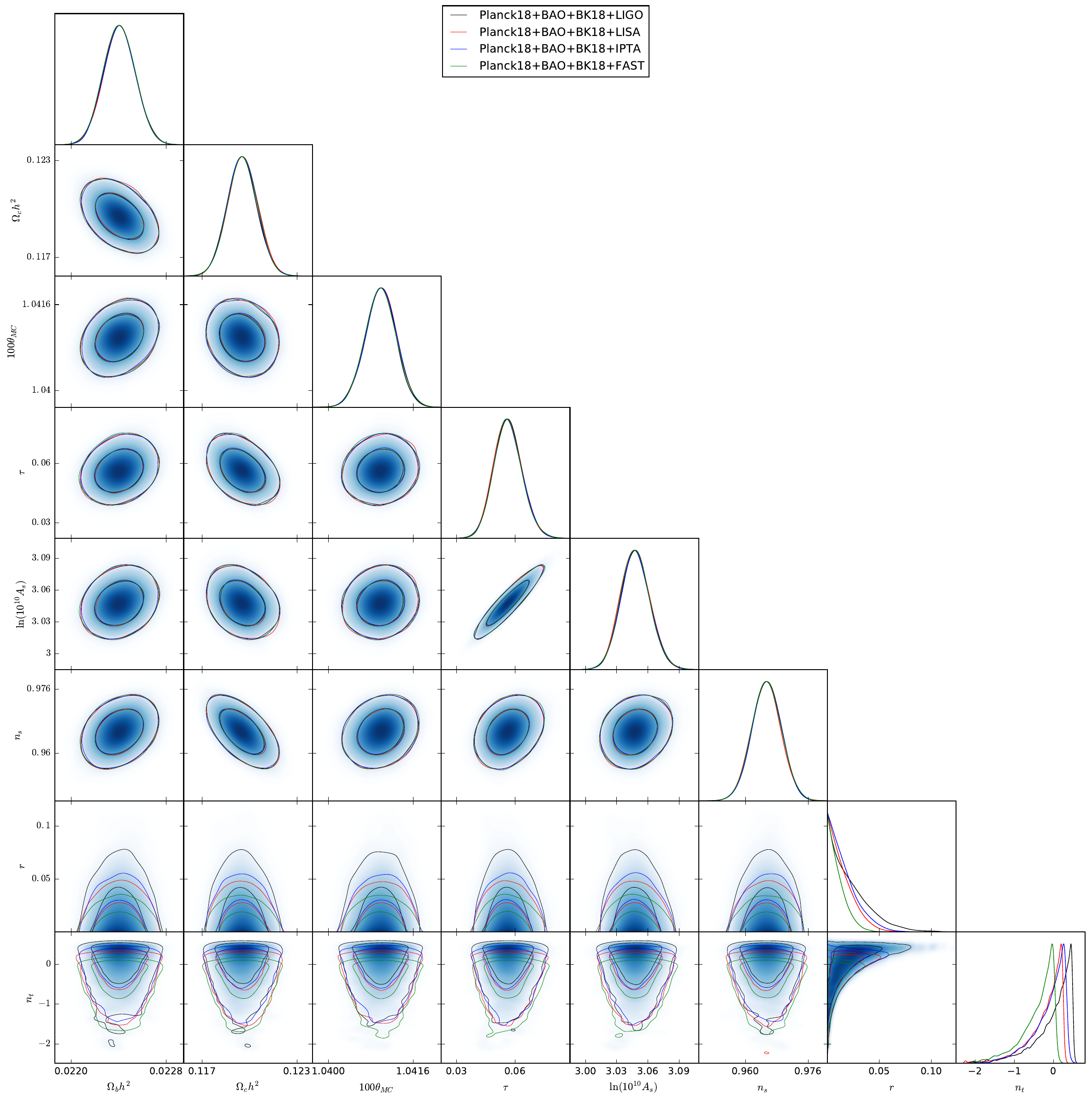}
\caption{The contour plots and the likelihood distributions for the cosmological parameters in the $\Lambda$CDM+$r$+$n_t$ model at the $68\%$ and $95\%$ CL from the combinations of Planck18+BAO+BK18+LIGO, Planck18+BAO+BK18+LISA, Planck18+BAO+BK18+IPTA and Planck18+BAO+BK18+FAST, respectively.}
\label{fig:3}
\end{figure*}

We use the publicly available codes Cosmomc \cite{Lewis:2002ah} to constrain the primordial gravitational waves. In the standard $\Lambda$CDM model, the six parameters are the baryon density parameter $\Omega_b h^2$, the cold dark matter density $\Omega_c h^2$, the angular size of the horizon at the last scattering surface $\theta_\text{MC}$, the optical depth $\tau$, the scalar amplitude $A_s$ and the scalar spectral index $n_s$. We extend this model by adding the tensor-to-scalar ratio $r$ and the tensor spectral index $n_t$, and consider these eight parameters as fully free parameters. 
We use the observations of SGWB from LIGO, LISA, IPTA and FAST to help measure $r$ and $n_t$ on the bases of CMB observations. Only CMB observations can not constrain $n_t$.
Our numerical results are given in Table.~\ref{table:1}, Fig.~\ref{fig:2} and Fig.~\ref{fig:3}.

\begin{table*}
\newcommand{\tabincell}[2]{\begin{tabular}{@{}#1@{}}#2\end{tabular}}
  \centering
  \begin{tabular}{ c | c| c| c| c}
  \hline
  \hline
  Parameters & \tabincell{c}{Planck18+BAO\\+BK18+LIGO} & \tabincell{c}{Planck18+BAO\\+BK18+LISA} & \tabincell{c}{Planck18+BAO\\+BK18+IPTA} & \tabincell{c}{Planck18+BAO\\+BK18+FAST}\\
  \hline
  $\Omega_bh^2$ &   $0.02241\pm0.00013$ & $0.02241\pm0.00013$&$0.02241\pm0.00013$&$0.02241\pm0.00013$\\
  $\Omega_ch^2$ &    $0.11952^{+0.00093}_{-0.00094}$ &$0.11957^{+0.00095}_{-0.00094}$ &$0.11954\pm{0.00093}$&$0.11955^{+0.00094}_{-0.00093}$\\
  $100\theta_{\mathrm{MC}}$ &   $1.04098\pm0.00029$ &$1.04099\pm0.00029$&$1.04099\pm0.00029$&$1.04099\pm0.00029$\\
  $\tau$ &   $0.0564^{+0.0071}_{-0.0073}$ &$0.0562^{+0.0070}_{-0.0078}$&$0.0564^{+0.0070}_{-0.0078}$&$0.0564^{+0.0069}_{-0.0077}$\\
  $\ln\(10^{10}A_s\)$  & $3.048\pm0.014$  &$3.048^{+0.014}_{-0.015}$&$3.048\pm0.014$&$3.048^{+0.014}_{-0.015}$\\
  $n_s$ &   $0.9654\pm0.0038$  &$0.9653^{+0.0037}_{-0.0038}$&$0.9654^{+0.0038}_{-0.0037}$&$0.9654\pm0.0038$\\
  $r_{0.05}$ ($95\%$ CL) &   $<0.061$ &$<0.038$&$<0.044$&$<0.028$\\
  $n_{t}$ ($95\%$ CL) &   $-0.02^{+0.72}_{-1.10}$&$-0.22^{+0.65}_{-0.95}$&$-0.13^{+0.65}_{-0.94}$&$-0.41^{+0.64}_{-0.96}$\\
  \hline
  \hline
  \end{tabular}
  \caption{The $68\%$ limits on the cosmological parameters from the combinations of Planck18+BAO+BK18+LIGO, Planck18+BAO+BK18+LISA, Planck18+BAO+BK18+IPTA and Planck18+BAO+BK18+FAST in the 
  $\Lambda$CDM+$r$+$n_t$ model.}
  \label{table:1}
\end{table*}

In the $\Lambda$CDM+$r$+$n_t$ model, the standard six parameters change slightly from SGWB observations.  
The constraint on scalar spectral index is 
\m
n_s=0.9654\pm0.0038\quad(68\% \ \mathrm{C.L.}),
\n
from Planck18+BAO+BK18+LIGO datasets; 
\m
n_s=0.9653^{+0.0037}_{-0.0038}\quad(68\% \ \mathrm{C.L.}),
\n
from Planck18+BAO+BK18+LISA datasets; 
\m
n_s=0.9654^{+0.0038}_{-0.0037}\quad(68\% \ \mathrm{C.L.}),
\n
from Planck18+BAO+BK18+IPTA datasets; 
\m
n_s=0.9654\pm0.0038\quad(68\% \ \mathrm{C.L.}),
\n
from Planck18+BAO+BK18+FAST datasets; 
These numerical results reveal that CMB observations are sensitive to the standard six parameters. Gravitational waves experiments can not measure them effectually so far. 

However, the constraint on tensor-to-scalar ratio and tensor spectral index are improved obviously from SGWB observations. The constraint on parameter $r$ and $n_t$ from Planck18+BAO+BK18+LIGO datasets is given by
\m
r &<& 0.061\quad(95\% \ \mathrm{C.L.}),\\
n_t &=&  -0.02^{+0.72}_{-1.10}\quad(95\%\ \mathrm{C.L.}).
\n
The constraint on parameter $r$ and $n_t$ becomes
\m
r &<& 0.038\quad(95\% \ \mathrm{C.L.}),\\
n_t &=&  -0.22^{+0.65}_{-0.95}\quad(95\%\ \mathrm{C.L.}),
\n
from Planck18+BAO+BK18+LISA datasets;
\m
r &<& 0.044\quad(95\% \ \mathrm{C.L.}),\\
n_t &=&  -0.13^{+0.65}_{-0.94}\quad(95\%\ \mathrm{C.L.}),
\n
from Planck18+BAO+BK18+IPTA datasets;
\m
r &<& 0.028\quad(95\% \ \mathrm{C.L.}),\\
n_t &=&  -0.41^{+0.64}_{-0.96}\quad(95\%\ \mathrm{C.L.}),
\n
from Planck18+BAO+BK18+FAST datasets; FAST projects present better constraints on the tensor-to-scalar ratio and tensor spectral index which is obvious in Table.~\ref{table:1} and Fig.~\ref{fig:2}.

\begin{figure*}
\centering
\includegraphics[width=15.8cm]{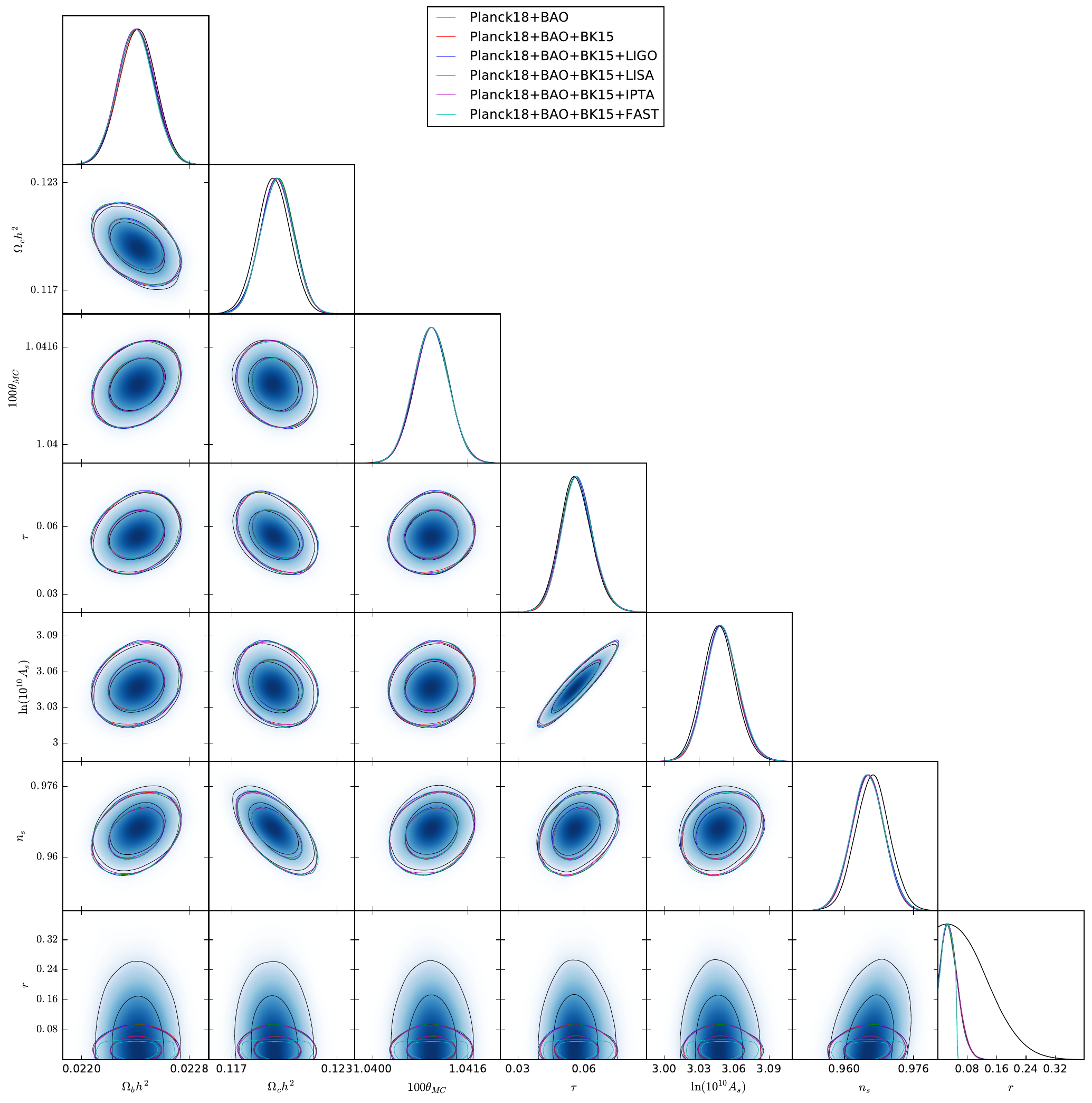}
\caption{The contour plots and the likelihood distributions for the cosmological parameters in the $\Lambda$CDM+$r$ model at the $68\%$ and $95\%$ CL from the combinations of Planck18+BAO, Planck18+BAO+BK15, Planck18+BAO+BK15+LIGO, Planck18+BAO+BK15+LISA, Planck18+BAO+BK15+IPTA and Planck18+BAO+BK15+FAST, respectively.}
\label{fig:4}
\end{figure*}

We also consider the $\Lambda$CDM+$r$ model which set $n_t=0$ and present our numerical results from the combinations of Planck18+BAO, Planck18+BAO+BK15 and Planck18+BAO+BK15+SGWB in Table.~\ref{table:2} and Fig.~\ref{fig:4}. FAST projects present better constraint on the tensor-to-scalar ratio which is obvious from Eq.~(\ref{FAST}).

\begin{table*}
\newcommand{\tabincell}[2]{\begin{tabular}{@{}#1@{}}#2\end{tabular}}
  \centering
  \begin{tabular}{c | c | c | c| c| c| c}
  \hline
  \hline
  Parameters & \tabincell{c}{Planck18+BAO}& \tabincell{c}{Planck18+BAO\\+BK15}  & \tabincell{c}{Planck18+BAO\\+BK15+LIGO} & \tabincell{c}{Planck18+BAO\\+BK15+LISA} & \tabincell{c}{Planck18+BAO\\+BK15+IPTA} & \tabincell{c}{Planck18+BAO\\+BK15+FAST}\\
  \hline
  $\Omega_bh^2$ &  $0.02241\pm0.00013$ & $0.02240\pm0.00013$ & $0.02240^{+0.00014}_{-0.00013}$ & $0.02240\pm0.00013$ & $0.02241\pm0.00013$ & $0.02240\pm0.00013$\\
  $\Omega_ch^2$ &   $0.11938\pm0.00093$ & $0.11957\pm0.00094$ & $0.11958\pm0.00096$ & $0.11959^{+0.00094}_{-0.00095}$ & $0.11957\pm0.00093$  & $0.11958\pm0.00094$\\
  $100\theta_{\mathrm{MC}}$ &  $1.04099\pm0.00029$ & $1.04099\pm0.00029$ & $1.04099\pm0.00029$ & $1.04099\pm0.00029$ & $1.04099\pm0.00029$ & $1.04099\pm0.00029$\\
  $\tau$ &  $0.0563^{+0.0070}_{-0.0076}$ & $0.0568^{+0.0069}_{-0.0076}$ & $0.0567^{+0.0069}_{-0.0078}$ & $0.0567^{+0.0069}_{-0.0076}$ & $0.0568^{+0.0069}_{-0.0077}$& $0.0569^{+0.0070}_{-0.0078}$\\
  $\ln\(10^{10}A_s\)$  & $3.047\pm0.014$ & $3.049\pm0.014$ &$3.049^{+0.014}_{-0.015}$ &$3.049^{+0.014}_{-0.015}$ & $3.049\pm0.014$  &$3.049^{+0.014}_{-0.015}$\\
  $n_s$ &  $0.9665\pm0.0038$ & $0.9655\pm0.0037$ & $0.9654\pm0.0039$  & $0.9654^{+0.0037}_{-0.0038}$ & $0.9655\pm0.0037$ & $0.9654\pm0.0038$\\
  $r_{0.05}$ ($95\%$ CL) &  $<0.212 $ & $<0.075$  & $<0.076$ & $<0.075$ & $<0.075$ & $<0.049$\\
  \hline
  \hline
  \end{tabular}
  \caption{The $68\%$ limits on the cosmological parameters in the $\Lambda$CDM+$r$ model from the combinations of Planck18+BAO, Planck18+BAO+BK15, Planck18+BAO+BK15+LIGO, Planck18+BAO+BK15+LISA, Planck18+BAO+BK15+IPTA and Planck18+BAO+BK15+FAST.}
  \label{table:2}
\end{table*}

\section{summary}
In summary, we constrain the primordial gravitational waves from CMB and SGWB observations. We not only combine LIGO observations with CMB to measure primordial gravitational waves, but also forecast the potential abilities of the LISA detector and PTA projects. In the $\Lambda$CDM+$r$+$n_t$ model, the standard six parameters change slightly from SGWB observations. While the constraint on tensor-to-scalar ratio and tensor spectral index are improved obviously from SGWB observations. Combine Planck18+BAO+BK18 with FAST project, the constraint on tensor-to-scalar ratio and tensor spectral index become much better.

\noindent {\bf Acknowledgments}.
This work is supported by Natural Science Foundation of Shandong Province (grant No. ZR2021QA073) and Research Start-up Fund of QUST (grant No. 1203043003587).



\end{document}